\documentstyle[12pt,doublespace]{article}
\begin{document}
\title{\protect\large \bf Cluster diagonalization in systematically expanded
Hilbert spaces: application to models of correlated electrons}

\author{Jos\'{e} Riera \\
\protect\small \em Physics Division, Oak Ridge National Laboratory, \\
\protect\small \em Oak Ridge, Tennessee 37831-6373\\
\\
Elbio Dagotto\\
\protect\small \em Department of Physics, Center for Materials Research
and Technology, \\
\protect\small \em and Supercomputer Computations Research Institute,
Florida State University, \\
\protect\small \em Tallahassee, Florida 32306}
\date{}
\maketitle

\begin{abstract}
A method of cluster diagonalization in a systematically
expanded Hilbert space is described. We discuss some applications
of this procedure to
models of high-T$_c$ superconductors, like the $t - J$
and one and three bands Hubbard models in two dimensions. The results
obtained with this
method are compared against results obtained with other techniques dealing
with truncated Hilbert spaces. The relation between this
method of diagonalization in a reduced Hilbert space, and perturbation
theory and variational techniques is also discussed.

\end{abstract}


\newpage

\section{\protect\large \bf Introduction}

\hspace{2em}Since the discovery of high-T$_c$ superconductivity,$^1$
intensive theoretical work has been carried out to understand its properties.
Much of this effort was devoted to the analysis of two dimensional
electronic models,$^2$ in particular, the Hubbard$^3$ and
$t - J$ models.$^4$
In spite of their apparent simplicity, these models are very
difficult to study with analytical techniques. Actually,
there are no exact solutions of these models except in one dimension
(and even in this case, for the $t - J$ model only $J = 0$ and
$J = 2t$, i.e. the supersymmetric point, can be solved exactly).
In the parameter regime of interest for high-T$_c$
superconductivity,
these models can be regarded as strongly correlated electronic systems.
It is well known that most analytical methods, like Hartree-Fock $^5$
or RPA approximations, which are reliable for
weak coupling systems, have difficulties in dealing with
strongly correlated electrons.  The same problem arises in
approximations like slave boson mean-field techniques.$^6$
In particular, for the
$t - J$ model it is not easy to decouple the charge and spin degrees
of freedom.

One should also note that in mean field calculations it is
necessary to make assumptions about ground state properties.
Numerical methods, on the other hand, are not biased by
any ``a priori" assumptions, and they have provided much of the reliable
information available for these models, as well as a useful check
of predictions formulated by analytical approximations.
Among the most widely used numerical techniques are the Monte Carlo
algorithms.$^7$ In particular, the version that uses the Hubbard-
Stratonovich transformation has been applied to the Hubbard
model$^8$ and several important results have been obtained.
An alternative to Monte Carlo techniques is the Lanczos method $^9$
which essentially gives the ground state of a given model for a
finite lattice.
{}From the ground state, we can compute all static and dynamical properties,
and in this sense, we obtain a complete characterization of a model at
zero temperature except for finite size effects.$^{10}$

This technique has provided important information about models of
correlated electrons.
For example, let us consider
a very recent work$^{11}$ where the $t - J$ model at quarter filling
has been studied.
In this work, strong signals of $d_{x^2-y^2}$
superconductivity close to the phase separation
border were found. These indications come from the study of
pairing correlations, Meissner effect and flux quantization
in the $4 \times 4$ lattice.
At quarter filling there are an equal number of holes and electrons
and we expect that at this point the finite size effects are
small.
However, if we consider the region physically relevant for
high-T$_c$ superconductivity which is close to half filling (doping
fraction $x \cong 0.10$), the number of
holes is very small (2 for the $4 \times 4$ lattice ) and then we
would expect
a weak signal for hole superconductivity.
Actually, most of the exact diagonalization studies of the $t - J$ model
on this lattice, using realistic couplings, have not found any
indications of superconductivity.
Then, in order to study the phase diagram of the $t-J$ model,
its properties, and the relation superconductivity-phase separation
in the physically
relevant region, it appears to be necessary to analyze larger clusters.
However, the 32 sites lattice with 4 holes requires the
diagonalization of  a matrix of $\sim 2.25
\times 10^{10}$ states, which is unreachable with present-day computers.
Similar Hamiltonian matrix dimensions appear in many other situations.

In this paper we want to stress the need for developing new methods in
the context of diagonalization in a reduced basis set
in order to answer quantitatively the important
questions posed by models of high-T$_c$ superconductivity.
There are strong reasons why we should attempt to improve
diagonalization
schemes, rather than other approaches like
Monte Carlo methods. It is well known that
Monte Carlo simulations of fermionic models
present ``the minus sign problem",$^{12}$
which makes very difficult the study of these systems at the physically
interesting densities.
It is also well known that there are difficulties in the
analytical continuation procedure that is necessary to perform in
Monte Carlo calculations of dynamical
properties, and thus these techniques are not well developed.
The diagonalization procedures are free from the minus
sign problem, and as we mentioned at the beginning, all quantities
static and dynamical, can be computed from the ground state. Thus, it
is very important to extend these techniques to large clusters, and the
attempt discussed in this paper corresponds to a systematic expansion
of the Hilbert space.

\section{\protect\large \bf Systematic expansion of the basis set}

\hspace{2em}As it was described in the Introduction, the sizes of
the Hilbert
space necessary to study quantitatively problems relevant to high-$T_c$
superconductivity
are considerably larger than the dimensions that can
be reached with present computers (although the currently available
results for small clusters seem to be qualitatively reliable).
In this context, here we want to show that significant results can be
obtained by diagonalization of the Hamiltonian in a truncated
or reduced Hilbert space.$^{13}$ Some variations of this procedure have
been used for
many years in other fields such as chemical physics (see, for example,
Ref.$\:$14) where similar work has been recently discussed
by Wenzel and Wilson.$^{15}$

The method of diagonalization in a truncated basis is of course
justified only if a few coefficients $x_i$ of the ground state:
\begin{eqnarray}
\Psi_0 =  \sum_{\scriptstyle i } x_i \phi_i,
\end{eqnarray}
\noindent
have significant weight. In some cases, fairly accurate properties of
the ground state
can be reached even with a small fraction of the total Hilbert space.

There are two questions that must be addressed to implement the proposed
technique:

\begin{enumerate}
\item It is necessary to choose an appropriate basis $\{\phi_i\}$ according to
the
physics of the problem. For example,

\begin{itemize}
   \item real space $S^z$ representation for the $t-J$ model,
   \item momentum space representation for the one-band Hubbard model in
weak coupling.
\end{itemize}
\item The algorithm must be able to find the most significant states
that contribute to the ground state wave function.
\end{enumerate}

The outline of the method we have developed and present in this paper,
which we call ``systematic
expansion of the Hilbert space'' (SEHS), is the following:

\begin{enumerate}
\item start from as few as possible states chosen according to the
expected behavior of the system (knowing quantum numbers of the ground
state greatly simplifies the work);
\item at each step $i$ expand the Hilbert space by applying the
Hamiltonian, or at least part of it, to the current set of states;
\item diagonalize in the new enlarged Hilbert space using the Lanczos
method;
\item retain the states with the largest weight, such that the
dimension of the Hilbert space is $N_i = \lambda N_{i-1}$, $1 < \lambda
\leq 2$ (``slow growth'' approach);
\item go back to step 2 until convergence in the physical quantities
is achieved,
or until the largest available dimension in the computer is reached.
\end{enumerate}

In an ideal situation, the states chosen at the starting point should
correspond to those
that carry most of the weight in the exact ground state. For some sets of
parameters (couplings, densities)
it is possible to guess these states.
However, different sets of parameters may have different behaviors,
and usually it is not possible to predict at which point the crossover
between them will occur. For example, in the $t-J$ model, for $J \gg t$ the
holes are bound together in pairs, so we can take as starting point
states where the holes are in nearest neighbor sites. On the
other hand, for $J \ll t$ the holes are not bound and move
around independently of each other and then it is not correct to
take the same states as before as the initial state for the iterations.
In this situation, the ``pruning" of the Hilbert space retaining
the most weighted states as indicated in point 4 is essential to
improve or correct the initial starting set of states. As we
discuss below, this procedure effectively works as a systematic method to
obtain and improve variational states. Moreover, it allows the
dimension of the Hilbert space to grow at a slow rate and the
behavior of the energy results smoother than in the case of the
straight application of the Hamiltonian. Below we will apply the
proposed
method to several cases relevant to theories of high-T$_c$ superconductors.

\section{\protect\large \bf Study of the $t-J$ model}

\hspace{2em}Let us apply the SEHS method
to the $t-J$ model$^4$ which is
defined by the Hamiltonian:

\begin{eqnarray}
H = - t \sum_{\scriptstyle <i j>, \sigma}
(\tilde{c}^{\dagger}_{i,\sigma}\tilde{c}_{j,\sigma} +
\tilde{c}^{\dagger}_{j,\sigma}\tilde{c}_{i,\sigma})
+ J \sum_{ <i j>}
({\bf S}_{i}\cdot {\bf S}_{j} - \frac{1}{4} n_{i} n_{j}),
\end{eqnarray}

\noindent
where the notation is standard. The first term describes the
hopping of holes or kinetic energy, while the second one corresponds to the
antiferromagnetic Heisenberg interaction. In this model the size of
the Hilbert space grows roughly as $3^{N_s}$, where $N_s$ is the number of
sites of the lattice, after taking into account the constraint of no
double occupancy.
In two dimensions (2D), this model has been studied at all fillings on
the $4 \times 4$ cluster.$^{10}$
Up to 2 holes, clusters of up to 26 sites have also been considered.
$^{16}$

First, let us briefly discuss the application of this
method
to the two dimensional $t-J_z$ model$^{17}$ which is obtained from the $t-J$
model by eliminating the spin exchange term in the Heisenberg
interaction.
Consider the case of one hole. In the limit of $J_z/t \gg
1$, the ground state of this model consists of a state in
which the hole is located at an arbitrary site surrounded by
an otherwise perfect N\'{e}el state.
In this limit the dimension of the Hilbert space
needed to get the physics of the problem is just equal to one
(plus all states translationally equivalent).
Now, as $J_z/t$ is reduced to the most interesting region, i.e. $J_z/t
\leq 1$, the hole gains kinetic energy at the expense of magnetic
energy and starts to move away from its initial position. As the hole
hops, it leaves behind a trail of overturned spins called a
``string".$^{18}$
As $J_z/t$ is
lowered, one must take into account longer and longer strings.
However the important string excitations are still of finite length,
and then in this case it is enough to keep a fraction of the total Hilbert
space
to describe it.
For this model, the application of the Hamiltonian, i.e. the hopping
term, to expand the Hilbert
space at each step has a direct physical meaning.
As we have shown in a previous paper, $^{17}$ it is possible to converge
to the ground state energy with several digits of accuracy
by retaining a small fraction of the full Hilbert space. As an example,
in Table I, the energy of the system for two holes is shown
for a cluster of 50 sites and $J_z/t = 0.3$ as a function of the
dimension of the Hilbert space. It is clear that the new technique works
very
well in this case. For more details see Ref. 17.

Let us now consider the $t - J$ model with the full Heisenberg
interaction (Eq. (2)). In this case, even in the absence of
holes, the ground state is characterized by the presence of
spin wave excitations that reduce the antiferromagnetic order from
its N\'{e}el (classical) value. Thus, in principle, we not only need to
physically
describe the modification of the spin background in the vicinity
of the holes, but also the spin exchanges that take place at arbitrary
distances from the holes which contribute significantly to the spin
background. This qualitative difference between the $t-J$ and $t-J_z$
models
 can be detected by measuring
the distribution of weights $S(x)$ defined as the sum of the
weights $\mid x_i \mid ^2$ belonging to the interval $\left[ x, x +
\Delta \right]$. In Fig.1, we show $S(x)$ in the exact ground
state of the $4 \times 4$ lattice with two holes at $J_z = 0.6$ and $J=0.6$
(in general we take t=1), for the $t - J_z$ (Fig.1a) and $t - J$
(Fig.1b) models, respectively. It can be seen that in the latter, there is more
weight for very small absolute values of the coefficients $x_i$
of the ground state $\Psi_0$ (Eq. (1)).

Let us start the expansion of the Hilbert space from the same sets of
states considered for the $t-J_z$ model. At each step, the Hilbert
space is expanded by the application of both the hopping term and
the spin exchange term of the Heisenberg interaction.$^{19}$
In the language of perturbation theory, this is like a double expansion
around the Ising limit ($t-J_z$) with static holes, namely one or two
holes in an otherwise perfect N\'{e}el state. The expansion
with the spin exchange term of the Heisenberg interaction could be
regarded as a perturbation in the spin anisotropic parameter.
In Figs. 2-5, we show results for the $4 \times 4$ lattice .
These can be compared with results for the
exact ground state which can be easily computed.
In Fig. 2, the energy is shown as a function of the
dimension of the basis set, for two holes at $J = 0.2$. The
energies obtained with the ``truncation'' procedure (dot-dashed line)
are much better than the energies obtained without it (dashed line)
namely diagonalizing at step 3 of the method, but without truncating in
step 4. As explained before, this improvement helps in discarding states
with very small weight.
Finally, both
are much better than the energies obtained at each iteration
of the conventional Lanczos algorithm (full line).
In Fig. 3, the overlap between the variational wave functions
in the truncated Hilbert space with the exact ground state are
shown for both procedures with (dot-dashed line) and without
(dashed line) the elimination of the less weighted states or
``truncation''.
In Fig. 4, the evolution of the
hole-hole correlations at the {\em maximum} distance in this lattice
is shown as a function of the dimension of the Hilbert space. It can be
seen that the convergence with the ``truncation'' procedure is much faster
than without it, even for correlation functions. The notation in these
figures is the
same as for Fig. 2. A similar behavior was also obtained for the
spin-spin correlation at the maximum distance.

Finally, to complete the preliminary study on the
$4 \times 4$ lattice, we show in Fig. 5 the energies obtained
with the full basis set expansion  procedure starting from the
N\'{e}el state (curve labeled 0); from the N\'{e}el state and all the
states obtained from it by one spin exchange (curve labeled 1);
from the N\'{e}el state and all the states obtained from it by two
spin exchanges (curve labeled 2); and so on. The energies at the
beginning of each set correspond to the variational states
discussed in Ref. 20. We see that the energies
obtained with the new method starting from the N\'{e}el state are
considerably better,
even for a very small number of iterations, than those corresponding
to Dagotto-Schrieffer's variational states. As a conclusion, even though we
cannot
reach the ground state as accurately as we did for the $t-J_z$ model, we still
can obtain a very good variational state compared with other
states discussed in the literature for finite lattices.

Now let us discuss clusters that cannot be studied with the conventional
Lanczos approach for lack of enough memory in present-day computers.
We will show results obtained for the
$t - J$ model on the $6 \times 6$ lattice with two holes, and $J = 0.4$.
The dimension of the Hilbert space is, in this case, $2.55 \times 10^9$ states
using translational and spin reversal symmetries.
In Fig. 6, the energy is plotted as a function of the dimension of
the Hilbert space (in a logarithmic scale).
With a full line we show the energies obtained at each step of the conventional
Lanczos algorithm, while with a dashed line we plot the energies obtained
expanding the Hilbert space by applying the Hamiltonian, and at each step
diagonalizing in the enlarged space using the Lanczos method, i.e.
steps 2 and 3 of
the method described
above. Finally, with circles and diamonds, we show the points
obtained by retaining the most weighted states, i.e. step 4 of our
method. The long-dashed line in zig-zag shows the order in which
every point is obtained starting with the circle at the top.
It is clear that a better convergence is achieved with the full
procedure of the SEHS method. After reaching the maximum
dimension that can be handled with the available computer, it is also
possible to use an extrapolation procedure to extract results at the dimension
of the total Hilbert space, but we have not attempted such an analysis
in the present paper.
(The energy for this particular system has been estimated
with a Green's Function Monte Carlo technique$^{21}$ to be
near $\sim -20.0$.) In principle, one should also compute other
physical quantities of interest at each coupling,
and then also extrapolate them to the full dimension.

Presumably, we can attribute the slow convergence of the ground state
energy with the size of the Hilbert space to the highly nontrivial (and
fluctuating) spin-1/2
background. Then, the convergence is not going to deteriorate if we
put more  holes on
the lattice. On the other hand, Monte Carlo algorithms typically encounter
increasingly severe problems as the number of holes is increased, at
least if one remains close to half-filling.
The number of off-diagonal transitions for
both the hopping (dashed line) and the exchange (full line)
parts of the Hamiltonian as a function of the number of states
included in the basis set can be computed
at each step.
The result is that successive sets generated
during the process of enlargement of the Hilbert space are
increasingly more interacting, i.e. the Hamiltonian matrix becomes
more dense (See, for example, Fig. 11 in Ref. 13.)

\section{\protect\large \bf Application to the one-band Hubbard model}

\hspace{2em}The one-band Hubbard model is defined
by the Hamiltonian:

\begin{eqnarray}
H = - t \sum_{\scriptstyle <i j>, \sigma}
(c^{\dagger}_{i,\sigma}c_{j,\sigma} +
c^{\dagger}_{j,\sigma}c_{i,\sigma})
+ U \sum_{i}  n_{i,\uparrow}n_{i,\downarrow},
\end{eqnarray}

\noindent where the notation is standard.
The size of the Hilbert space grows
as $4^{N_s}$, and thus it is  even more difficult to study than the $t-J$
model from a numerical point of view.
In this case, the largest lattice considered in the literature
is the $4 \times 4$ lattice
for all dopings.$^{10,22}$
In momentum space, the Hamiltonian of the Hubbard model takes the
form:

\begin{eqnarray}
H = \sum_{\scriptstyle {\bf k }, \sigma} \epsilon ({\bf k })
c^{\dagger}_{ {\bf k } ,\sigma} c_{ {\bf k } ,\sigma} +
+ U \sum_{ {\bf k_1,k_2,k_3}}  c^{\dagger}_{ {\bf k_1 },\uparrow}
c_{ {\bf k_2},\uparrow} c^{\dagger}_{ {\bf k_3},\downarrow}
c_{ {\bf k_1-k_2+k_3},\downarrow},
\end{eqnarray}

\noindent
where each {\bf k } runs over the Brillouin zone. The single
particle energies are given by $\epsilon ( {\bf k }) = - 2 t
(cos( k_x ) + cos(k_y ))$.
In the absence
of Coulomb repulsion, the model reduces to a tight binding
model which is easily solved. The total energy is the sum of
the single particle energies for all the momentum ${\bf k }$
up to the Fermi surface. Here, we have to distinguish between
two cases: the closed shell, in which the last shell is completely
occupied; and the open shell in which the last shell is partially
occupied. In the former case the ground state is not degenerate
while in the latter the degeneracy can be very large.
In the following,
we concentrate on the $6 \times 6$ cluster with  18 (9$\uparrow$ and
9$\downarrow$) and 26 (13$\uparrow$ and 13$\downarrow$) electrons
which correspond to $closed$ shell situations.
The dimensions of the Hilbert space for some closed shell cases
in this cluster are: for 10 electrons, $3.95 \times 10^{9}$; for 18 electrons,
$2.46 \times 10^{14}$; and for
26 electrons,  $1.48 \times 10^{17}$ well beyond the reach of
techniques that fully diagonalize the full Hilbert space of the problem.

For the closed shell situations, our initial Hilbert space consists
of only one state, which is the ground state of the $U = 0$ case
(remember that we are working in momentum space).
The Hilbert space is expanded by applications of the second term
of Eq. (4), which contains the off diagonal transitions.
These terms create and annihilate pairs of
electrons in such a way that the total momentum is conserved.
In some other approaches the Hamiltonian is expanded through the
creation of single pair electron-hole excitations$^{23}$ but then
the total momentum is not conserved.
In the spirit of the general
procedure outlined in Section 2, we expand the Hilbert space by applying
the whole second term of Eq. (4). (Another possibility, which we have not
yet fully explored, is to expand the Hilbert space by taking only transitions
between the shells at both sides of the Fermi level, and then increase
successively the number of shells involved.)
The expansion of the Hilbert space by application of the Coulomb
term could also be considered as a weak-coupling perturbation
expansion in
a parameter which is proportional to $U$, but unlike other
perturbation schemes,$^{24}$ our procedure remains variational
in the sense that the energy is always an upper bound to the
exact ground state energy.$^{25}$

In Figs. 7 and 8, we show the convergence of the energy as a
function of the dimension of the Hilbert space for 18 and 26
electrons respectively, and for several values of $U$.
The energies are measured in units of $t$ as usual, and they
have been shifted in order to fit them into the same
plot and in order to compare their convergence.
It can be observed that the convergence is faster the fewer the electrons, and
as expected, the convergence is faster for smaller values of $U$.
For example, for the case of 26 electrons, for $U = 2$ we obtain
a value of -47.907, in good agreement with the Monte Carlo estimate
$^{26}$ of -47.87$\pm$0.05, i.e. the new technique reaches the same
accuracy as Monte Carlo methods.

The most important features in these plots are the presence of
discontinuities in the derivative of the energy,
and a ``wrong" concavity of the curves
(compared for example with the curvature in Figs. 5 and 6 of the
$t-J$ model). We do not have an explanation for this behavior,
although perhaps the long-range nature of the Coulomb interaction in
momentum space may matter.
The wrong curvature of the plots makes it difficult to assess
the convergence of the energy and to perform an extrapolation
procedure.
The points at which there are discontinuities in the derivative
are the points obtained by successive application of the
Hamiltonian starting from the initial state. All the other points
are obtained by pruning these
Hilbert spaces, and by applying the Hamiltonian to the reduced
spaces.
The somewhat strange behavior of the energy vs. the dimension of
the Hilbert space is an artifact of the momentum representation
chosen, and perhaps a manifestation of the shell structure
of the tight binding limit.

In the interval considered, i.e. $U \leq 4$, we found that the convergence
of the energies
obtained by working in the momentum representation is much faster
than the one obtained by working in real space. Presumably, the
opposite is true for larger values of $U$.

Finally, in Table II we provide comparisons of our estimates
with the results obtained using Quantum
Monte Carlo techniques,$^{26,27}$ as well as the results
obtained with a stochastic implementation of the modified Jacobi
method$^{28}$ also referred to as ``stochastic diagonalization'' (SD).

To obtain the results quoted in this Table, $N_R \sim 2 \times 10^4 $
important states were included in the SD calculation and a CPU
time of $\sim 10^4 $ seconds (for the $4 \times 4$ lattice) was required.
This CPU time is also what is required by our method for
$N_h \sim 10^6$. However, as reported in Ref. 28, and as it can be
seen in Table II, the energy is not yet converged and presumably $N_R$
has to be increased by a factor of $\sim 10$ in order to obtain the
same accuracy as our results. This translates to a factor of $\sim 100$
in the total CPU time, since in the SD algorithm the CPU time grows
quadratically with $N_R$. Besides, from the results reported in Ref.
28, it is also evident that for the SD method the convergence is
more difficult for larger values of the Coulomb repulsion.
In summary, it seems that at least in its current implementation, the
SD method is more expensive than the SEHS method reported in this
paper for a given accuracy.

\section{\protect\large \bf Application to the three-band Hubbard model}

\hspace{2em}Finally, and for completeness,
we briefly consider the three-band Hubbard model which contains
the Coulomb on-site repulsion for both the copper and
oxygen sites ($U_d$ and $U_p$ respectively), the energies of each ion
($e_d$ and $e_p$ for copper and oxygens ions), and a Coulomb
repulsion between copper and oxygens ions, $V$.$^{29}$
We study the $\sqrt{8} \times \sqrt{8}$ lattice (24 sites between
oxygens and
coppers) with
two doped holes (10 fermions), and the following set of parameters:
$U_d = 7$, $U_p = 0$, $e_p - e_d = 1.5$ and $V = 3$. As the initial
basis set, we took all the states with all the Cu sites having single
occupancy, and the remaining two holes located in O sites (also single
occupied). This is a good starting Hilbert space for the case
$V = 0$, but as the algorithm itself has shown it is not appropriate
for all values of the parameters.

In Fig. 9, we show the results obtained using the
Hilbert space expansion procedure. The dashed lines show the order
in which these points were obtained starting from the circle at the
top right in the same way as was explained in Fig. 6. In Fig. 10,
the best points
in the set of results shown in Fig. 9 are plotted with circles. In a
second stage, once we have reached $\sim 10^6$ states, we go
all the way back (points indicated with full
diamonds), finding that the initial guess was not appropriate (i.e. the
states with the highest weights were not those used in the starting Ansatz),
and then we increase the dimension of the basis set again
(empty squares). It can be seen that this last set of points
behaves very smoothly and the final part of the curve is fairly
flat indicating a reasonable convergence. From this set of
states, in principle, we could compute all quantities of interest and
eventually
extrapolate them to the full Hilbert space.
However, one should also notice that in this case the largest dimension that
we have considered ($\sim 2.5 \times 10^6$) is ``only" two orders of magnitude
smaller than the dimension of the full Hilbert space, and probably that
is the reason for the good convergence of the results.

In Fig. 11 we compare the energies for  $V = 0$ and $V = 3$ as a
function of the dimension of the Hilbert space. The energies have been
shifted for the sake of comparison. It can be seen that the
convergence is better for the $V = 3$ case. For $V = 0$, following the
Zhang-Rice construction,$^4$ one can map this model to the one-band
$  t - J $ model. It is then reasonable to assume that, as in this
model, the spin background is responsible for the slow convergence.

The same pattern of convergence was also found for the other set of
parameters we have studied: $U_p = 3$, $e_p - e_d = 4$,
and $V = 0$, $V = 3$, and the same value of $U_d = 7$. In this
case, for $V = 3$ the convergence
is faster than for $V = 0$, reflecting the fact that it is easier for
the algorithm to find the most relevant states which contains double
occupied Cu sites.

Finally, we show in Fig. 12 the spin-spin correlation
at the maximum distance on the lattice, and the density of holes in
Cu sites as a function of the dimension of the Hilbert space for
the set of parameters $U_d = 7$, $U_p = 0$, $e_p - e_d = 1.5$ and
$V = 3$. These curves
indicate also a reasonable convergence. For this set of parameters
we obtain $n_{Cu} = 0.555$, while for $U_p = 3$, $e_p - e_d = 4$,
$n_{Cu} = 1.088$, indicating the presence of two different regimes
for large $V$. This result might be relevant to some speculation
regarding the nature of pairing and phase separation in $Cu-O$
planes.$^{30}$


In any case, it is quite encouraging to observe that the new technique
may work well in the realistic (and complicated) case of the three-band
Hubbard model.

\section{\protect\large \bf Discussion and conclusions}

\hspace{2em}

The procedure described in this paper can be regarded as a method to generate
and/or
improve variational wave functions. In the first place, it should be noted
that since no approximations are done on the Hamiltonian, and since
we work in a reduced Hilbert space, the energies obtained with
this procedure are rigorous upper bounds to the exact ground
state energies. The application to the $t-J$ model is one example
in which the initial set of states is ``corrected" by this algorithm.
In this case, a direct comparison with a variational
state was also given (see also Ref. 19).
Another application in which the elimination at each step of the
least weighted states leads to an improvement or to a correction
of the initial guess is the case of the three-band Hubbard model.
In this case, the initial state depends on the parameters
that determine the $Cu$ or $O$ occupancy when the nearest
neighbor Coulomb repulsion is large enough. In general, we believe that
the technique is promising and may compete against more standard Lanczos
and
Quantum Monte Carlo methods, at least for some particular Hamiltonians
and parameters. A clear example is the $t-J_z$ model in which the new
method has provided the more accurate results reported in the literature
thus far.$^{17}$

For the systems where we cannot arrive at a good approximation for the
ground state due to the slow rate of convergence of the results (for example
the $t-J$ model seems to converge only logarithmically), one should
resort to some extrapolation procedure to the full Hilbert dimension.
In this sense, we are in the same situation as the zero temperature
(Green's function or random walk) Monte Carlo algorithms that
cannot reach convergence before the noise becomes very high.$^{21}$

Besides the possible applications of this reduced Hilbert space
approach as indicated above, there are other situations that can also
be studied with the SEHS method. One of them is the quarter-filled
$t-J$ model on the 26 sites lattice, which is interesting to
study in order to analyze the finite size dependence of the results obtained in
Ref. 11 in the context of superconductivity in the $t-J$ model.
The method can also be applied to
coupled planes $t-t_\perp-J$
model.$^{31}$ For this system, one could start from the best states
of the ground state of each plane separately and then expand the
basis set by application of the interplane hopping term of the
Hamiltonian. This is equivalent to an expansion around
$t_\perp / t = 0$.

Finally, we want to comment that there are other algorithms that also
deal with truncated Hilbert spaces besides the stochastic
diagonalization approach and the presently described technique.
In an already mentioned paper,$^{23}$ the Hubbard model was studied
in momentum space with a truncation technique using concepts
of renormalization group theory.
Another stochastic truncation method has recently
been developed for the $Z_2$ gauge model.$^{32}$
The computational effort of the SEHS method of systematic
expansion of the Hilbert space grows roughly linearly with $N_h$, and
currently $N_h \sim 10^6$ for present-day computers.
These other methods use a smaller
size of the basis set, but the CPU time grows
as ${N_h}^3$ for the methods of references [23] and [32], and
quadratically in $N_h$ for the stochastic diagonalization algorithm.

Summarizing, a new algorithm has been discussed that has several
of the advantages of the Lanczos approach (specially the possibility of
studying dynamical responses), but that can be applied to large
clusters. The method works remarkably well in some special cases,
while in general it is competitive with other more standard algorithms.

\section{\protect\large \bf Acknowledgements}

\hspace{2em}
We thank Adriana Moreo for providing the Monte Carlo results
used in this paper, and for useful conversations. E. D. thanks
the Office of Naval Research for its partial support under
grant ONR-N00014-93-1-0495. J. R. wishes to acknowledge the support
from High Performance Computations grant from Vanderbilt
University.
Most of the calculations were done using the Cray YMP at the
Supercomputer Computations Research Institute in Tallahassee,
Florida. The research was sponsored in part by the U. S. Department
of Energy under contract No. DE-AC05-84OR21400 managed by Martin
Marietta Energy Systems, Inc.

\newpage

\section{\protect\large \bf References}

\begin{enumerate}
\item J. G. Bednorz and K. M\"uller, {\em Z. Phys.} {\bf B 64}, 189 (1986).
\item P. W. Anderson, {\em Science} {\bf 235}, 1196 (1987).
\item J. Hubbard, {\em Proc. R. Soc. London, Ser.} {\bf A 276}, 238 (1963).
\item F. Zhang and T. M. Rice, {\em Phys. Rev.} {\bf B 37}, 3759 (1988).
\item See for example, J. A. Verg\'{e}s, E. Louis, P. S. Lomdahl, F.
Guinea and A. R. Bishop, {\em Phys. Rev. } {\bf B 43}, 4462 (1989), and
references therein.
\item G. Kotliar and A. E. Ruckenstein, Phys. Rev. Lett. {\bf 57}, 1362 (1986).
\item W. von der Linden, {\em Phys. Rep. } {\bf 220 }, 53 (1992).
\item A. Moreo, D. J. Scalapino, R. L. Sugar, S. R. White and N. E.
Bickers, {\em Phys. Rev.} {\bf B 41}, 2313 (1990); and references therein.
\item B. N. Parlett ,{\em  ``The symmetric eigenvalue problem"},
(Prentice Hall, 1980).
\item E. Dagotto, A. Moreo, F. Ortolani, D. Poilblanc
and J. Riera, {\em Phys. Rev. } {\bf B 45}, 10741 (1992).
\item E. Dagotto and J. Riera, {\em Phys. Rev. Lett.} {\bf 70},
682 (1993).
\item E. Y. Loh, et al., Phys. Rev. {\bf 41}, 9301 (1990).
\item An earlier discussion of this method was given in J. Riera, in
``Proceedings of the Mardi Gras '93 Conference on Concurrent Computing
in the Physical Sciences", World Scientific, 1993.
\item P. J. Knowles, { \em Chem. Phys. Letters} { \bf 155}, 513
(1989); P. J. Knowles and N. C. Hardy, { \em J. Chem. Phys.} { \bf 91},
2396 (1989).
\item W. Wenzel and K. G. Wilson, Phys. Rev. Lett. {\bf 69}, 800 (1992).
\item D. Poilblanc, J. Riera, and E. Dagotto, preprint, (1993).
\item J. Riera and E. Dagotto, {\em Phys. Rev. } {\bf B 47}, xxxxx (1993).
\item W. F. Brinkman and T. M. Rice, {\em Phys. Rev.} {\bf B 2}, 1324 (1970);
B. I. Schraiman and E. D. Siggia, {\em Phys. Rev. Lett.} {\bf 60},
740 (1988).
\item In a semi-analytical approach (S. Trugman, {\em Phys. Rev.} {\bf B 37},
1597 (1988); {\em Phys. Rev.} {\bf B 41}, 892 (1990)), the basis set was
expanded by
the application of the hopping term only (and the second neighbor double
hopping term present in the model considered by Trugman). See also J.
Inoue and
S. Maekawa, Prog. Theor. Phys., Suppl. {\bf 108}, 313 (1992).
Typically,
the Hilbert space was expanded to include a few hundred states. This is a very
small quantity compared with the $\sim 10^6$ one can reach with our
method, but Trugman's results are valid for the bulk limit. So, we obtain a
much better variational state but at the cost of limiting ourselves to
finite lattices.
\item E. Dagotto and J. R. Schrieffer, {\em Phys. Rev. }{\bf B 43},
8705 (1991).
\item M. Boninsegni and E. Manousakis, preprint (1992).
\item G. Fano, F. Ortolani and A. Parola, {\em Phys. Rev.} {\bf B 42}, 6878
(1990).
\item S. R. White, {\em Phys. Rev.} {\bf B 45}, 5752 (1992).
\item J. Gal\'{a}n and J. A. Verg\'{e}s, {\em Phys. Rev. } {\bf B 44},
10093 (1991).
\item  A numerical, but more conventional, weak-coupling perturbative study
on the $6 \times 6$ lattice was reported by B. Friedman,
{ \em Europhysics Letters} { \bf 14}, 495 (1991).
\item A. Moreo, private communication.
\item N. Furukawa and M. Imada,
{\em J. Phys. Soc. Jpn.} {\bf 61}, 3331 (1992).
\item H. De Raedt and W. von der Linden, {\em Phys. Rev.} {\bf B 45},
8787 (1992); H. de Raedt and M. Frick, {\em Phys. Rep.}, to appear. See
also
P. Prelovsek, and X. Zotos, preprint.
\item V. Emery, {\em Phys. Rev. Lett.} {\bf 58}, 2794 (1987)
\item C. Varma,
S. Schmitt-Rink and E. Abrahams, {\em Solid State Commun.} {\bf 62}, 681
(1987).
\item J. M. Wheatley, T. C. Hsu and P. W. Anderson, {\em Nature} {\bf 333},
121 (1988).
\item C. J. Hamer and J. Court, preprint (1992).

\end{enumerate}

\newpage

\noindent
{\bf Table I}

\vskip 0.8cm

\begin{tabular}{rr}         \hline\hline
${\rm H_D}$    &    $E_{2h}$   \\  \hline
234    &    -18.707940   \\
696    &    -18.882805   \\
6204    &    -19.026339   \\
18416    &    -19.052528   \\
52672    &    -19.066660   \\
106435    &    -19.074957   \\
212486    &    -19.079975   \\
673640    &    -19.083531   \\
980681    &    -19.084816   \\
1502829    &    -19.085503   \\
2249454    &    -19.085857    \\  \hline\hline
\end{tabular}

\noindent

\vskip 2cm

{\bf Table II}
\vskip 0.8cm

\begin{tabular}{llr}         \hline\hline
method     &    18 electrons   &    26 electrons   \\  \hline
QMC      &    -41.87$\pm$0.10     &      -41.98$\pm$0.15   \\
SEHS     &    -41.69    &       -41.49   \\
SD      &     -41.45    &      -40.77    \\  \hline\hline
\end{tabular}

\newpage
\centerline {\bf TABLE CAPTIONS}
\vskip 2truecm

\noindent
{\bf Table I}

\noindent
Energy $E_{2h}$ of two holes in the ${\rm t-J_z}$ model, as a function
of the size of the Hilbert space, ${\rm H_D}$, for a cluster of
50 sites, and coupling $J_z/t = 0.3$.

\vskip 2truecm

\noindent
{\bf Table II}

\noindent
Comparison between ground state energies (in units of $t$) obtained
with the present method (SEHS), Quantum Monte Carlo (QMC), and
Stochastic Diagonalization (SD), for the $6 \times 6$ lattice and
$U = 4$.

\newpage
\centerline {\bf FIGURE CAPTIONS}
\vskip 2truecm

\noindent
{\bf Figure 1}
\noindent
Distribution of weights S(x) a) for the $t -  J_z$ model, b) for
the $t -  J$ model on the $4 \times 4$ lattice with 2 holes and $J/t = 0.6$.
\vskip 1truecm

\noindent
{\bf Figure 2}
\noindent
Energy vs dimension of the Hilbert space for the $4 \times 4$
lattice with two holes, J = 0.4. The full curve corresponds to
the energies obtained at each step of the conventional Lanczos
iteration. The dot-dashed (dashed) corresponds to the procedure
indicated in Sec. 3 with (without) including step 4.
\vskip 1truecm

\noindent
{\bf Figure 3}
\noindent
Overlap between the exact ground state and the states
generated during the procedure of expansion of the Hilbert space.
The meaning of the curves are as in Fig. 2.
\vskip 1truecm

\noindent
{\bf Figure 4}
\noindent
Hole-hole correlations at the maximum distance on the $4 \times 4$
lattice. The meaning of the curves are the same as for Fig. 2.
\vskip 1truecm

\noindent
{\bf Figure 5}
\noindent
Expansion of the Hilbert space starting from different
initial basis sets for the $4 \times 4$ lattice with 2 holes and J=0.2.
\vskip 1truecm

\noindent
{\bf Figure 6}
Energy vs dimension of the Hilbert space for the $6 \times 6$
lattice, 2 holes, J=0.4.
\noindent
\vskip 1truecm

\noindent
{\bf Figure 7}
\noindent
Energy of the Hubbard model on the $6 \times 6$ lattice with
18 electrons vs dimension of the Hilbert space. The asterisk
indicate the Monte Carlo estimates.
\vskip 1truecm

\noindent
{\bf Figure 8}
\noindent
Energy of the Hubbard model on the $6 \times 6$ lattice with
26 electrons vs dimension of the Hilbert space. The asterisk
indicate the Monte Carlo estimates.
\vskip 1truecm

\noindent
{\bf Figure 9}
Energy of the three-band Hubbard model on the
8 cells square lattice as obtained by application of the SEHS
procedure.
\vskip 1truecm

\noindent
{\bf Figure 10}
Energy of the three-band Hubbard model on the
8 cells square lattice vs the dimension of the Hilbert space.
The open circle points correspond to the filled square points
of Fig. 12. After reaching $\sim 10^6$ states, we truncate
the Hilbert space in successive steps (diamonds), and then
we start a new expansion of the basis set (squares).
\vskip 1truecm

\noindent
{\bf Figure 11}
Energy of the three-band Hubbard model on the
8 cells square lattice vs the dimension of the Hilbert space
for different values of the intersite Coulomb repulsion $V$.
\vskip 1truecm

\noindent
{\bf Figure 12}
Spin-spin correlation at the maximum distance and density of
holes at Cu sites for the three-band Hubbard model on the
8 cells square lattice vs the dimension of the Hilbert space.
\vskip 1truecm

\end{document}